\documentclass[prb,showpacs,twocolumn,superscriptaddress]{revtex4}
\usepackage[latin2]{inputenc}
\usepackage{amsmath}
\usepackage{amssymb}
\usepackage{graphicx}
\usepackage{dcolumn}
\usepackage{epstopdf}

\def\ov{\overline}

\begin{document}

\title{Spin-orbit interaction and Dirac cones in $d$ orbital noble metal surface states}

\author{Ryan Requist}
\affiliation{
International School for Advanced Studies (SISSA), Via Bonomea 265, 34136 Trieste, Italy
}
\author{Polina M.~Sheverdyaeva}
\affiliation{
Istituto di Struttura della Materia, Consiglio Nazionale delle Ricerche, Trieste, Italy
}
\author{Paolo Moras}
\affiliation{
Istituto di Struttura della Materia, Consiglio Nazionale delle Ricerche, Trieste, Italy
}
\author{Sanjoy K.~Mahatha}
\affiliation{
International Centre for Theoretical Physics (ICTP), Strada Costiera 11, 34151 Trieste, Italy
}
\author{Carlo Carbone}
\affiliation{
Istituto di Struttura della Materia, Consiglio Nazionale delle Ricerche, Trieste, Italy
}
\author{Erio Tosatti}
\affiliation{
International School for Advanced Studies (SISSA), Via Bonomea 265, 34136 Trieste, Italy
}
\affiliation{
International Centre for Theoretical Physics (ICTP), Strada Costiera 11, 34151 Trieste, Italy
}
\affiliation{
Istituto Officina dei Materiali, Consiglio Nazionale delle Ricerche -- Democritos, Trieste, Italy
}

\date{\today}

\begin{abstract}

Band splittings, chiral spin polarization and topological surface states generated by spin-orbit interactions at crystal surfaces are receiving a lot of attention for their potential device applications as well as fascinating physical properties.  Most studies have focused on $sp$ states near the Fermi energy, which are relevant for transport and have long lifetimes.  Far less explored, though in principle stronger, are spin-orbit interaction effects within $d$ states, including those deep below the Fermi energy.  Here, we report a joint photoemission/\textit{ab initio} study of spin-orbit effects in the deep $d$ orbital surface states of a 24-layer Au film grown on Ag(111) and a 24-layer Ag film grown on Au(111), singling out a conical intersection (Dirac cone) between two surface states in a large surface-projected gap at the time-reversal symmetric $\overline{M}$ points.  Unlike the often isotropic dispersion at $\ov{\Gamma}$ point Dirac cones, the $\overline{M}$ point cones are strongly anisotropic. An effective $\mathbf{k}\cdot\mathbf{p}$ Hamiltonian is derived to describe the anisotropic band splitting and spin polarization near the Dirac cone.

\end{abstract}


\maketitle

\section{Introduction \label{sec:introduction}}

Spin-orbit coupling is a relativistic effect which significantly alters the electronic structure of materials containing heavy elements.  For instance, it lowers the degeneracy of non-relativistic bands at all wave vectors $\mathbf{k}$ where the double point group symmetry is lower than the point group symmetry \cite{kittel1963}.  The resulting band splittings are generally larger in $d$ bands, for which the atomic spin-orbit interaction $\xi \:\mathbf{L}\cdot\mathbf{S}$ is larger, than in $sp$ bands. In nonmagnetic crystals there always remains a two-fold Kramer's degeneracy which spin-orbit coupling cannot lift because it is protected by time-reversal symmetry.  In crystals with both time-reversal and inversion symmetry, this Kramer's degeneracy implies the spin degeneracy $\epsilon_{n\uparrow}(\mathbf{k})=\epsilon_{n\downarrow}(\mathbf{k})$.  Spin degeneracy can be lifted at interfaces, where inversion symmetry is obviously broken, by the Rashba-Bychkov effect \cite{rashba1960a,bychkov1984a,bychkov1984b}.  The Rashba-Bychkov effect results from the chiral coupling of momentum and spin according to
\begin{align}
H_{\rm SO}=\frac{\hbar^2}{4m^2c^2}\:(\mathbf{p}\times\mathbf{\sigma} )\cdot\mathbf{E}{,} \label{eq:Rashba}
\end{align}
where $\mathbf{E}=E_z \hat{\mathbf{z}}$ is the intrinsic electric field of the interface.  The lifting of spin degeneracy can be interpreted as a Zeeman splitting in a $\mathbf{k}$-dependent effective magnetic field $\mathbf{B}\sim \mathbf{E}\times\mathbf{k}$ or non-Abelian gauge field.

The Rashba-Bychkov effect has been studied for its relevance to potential spintronics applications \cite{zutic2004,awschalom2007}, e.g.~spin valves, spin pumps, spin field-effect transistors \cite{datta1990} and devices based on magnetic anisotropy.  Spin-orbit interactions have also received extra attention recently as the source of band inversion in 3d topological insulators \cite{fu2007a,moore2007,roy2009}.  Topological surface states, their Dirac cones and chiral spin polarization have been observed in $\mathbb{Z}_2$ topological insulators, e.g.~Bi and Sb compounds such as Bi$_{1-x}$Sb$_x$ and Bi$_2$Se$_3$ \cite{hsieh2008,xia2009,hsieh2009,chen2009,jozwiak2011}, and topological crystalline insulators \cite{fu2011}, e.g.~Pb and Sn compounds such as Pb$_{1-x}$Sn$_x$(Se,Te) \cite{dziawa2012,tanaka2012,xu2012,wojek2013}.  Metallic surface states satisfying a massless Dirac equation---precursors of topological insulator surface states---were first predicted to occur at the interface of two materials whose bands are mutually inverted \cite{volkov1985,fradkin1986,pankratov1987}, such as heterojunctions composed of Pb$_{1-x}$Sn$_x$Te, Pb$_{1-x}$Sn$_x$Se or Hg$_{1-x}$Cd$_{x}$Te.  The above examples illustrate the  importance of spin-orbit coupling in reduced geometries. 

The effects of Rashba-Bychkov spin splitting were first observed in the 2d electron gas in GaAs-AlGaAs heterojunctions \cite{stormer1983,stein1983}. When the spin-orbit interaction $H_{\rm SO}$ is added to a 2d free electron gas, it splits the otherwise spin-degenerate free-electron parabolas into two branches, forming a ``mexican hat'' surface of revolution, $E(\mathbf{k}) = \frac{\hbar^2|\mathbf{k}|^2}{2m^*} \pm \alpha_R \mathbf{k}$, with a conical intersection at the origin in $\mathbf{k}=(\mathbf{k_x},\mathbf{k_y}$) space, where $\alpha_R$ is the Rashba parameter.  Since the Shockley surface states which exist on close-packed noble metal surfaces also form a quasi-2d free electron gas and the surface explicitly breaks inversion symmetry, spin-orbit interactions can potentially lift spin degeneracy there too. Spin splitting was indeed observed by angle-resolved photoemission spectroscopy (ARPES) in the $sp$-derived $L$-gap surface states on Au(111) \cite{lashell1996}.  Spin splitting was also observed in $d$-derived surface states on W(110) \cite{rotenberg1998} and Mo(110) \cite{rotenberg1999} as well as its enhancement with the adsorption of monovalent atoms.  Even the small spin splitting of the Cu(111) $L$-gap state has recently been observed \cite{tamai2013}.  The dispersion of the $L$-gap surface states on Au(111) can be accurately fit by the above parabolic function with $m^*=0.255\: m_e$ and $\alpha_R=6.0\times 10^{-11}$~eV~m \cite{lashell1996,reinert2001,nicolay2001}.  Density functional theory (DFT) calculations in the local density approximation (LDA) reproduce the observed parabolic dispersion and spin splitting on Au(111) up to a rigid overall energy shift \cite{nicolay2001,forster2007}, while the calculated splitting on Ag(111) was too small to be resolved in experiment \cite{nicolay2001}.  The chiral spin polarization of the surface states on Au(111) \cite{hoesch2004,henk2004,muntwiler2004} and W(110)-(1$\times$1)H \cite{hochstrasser2002} was confirmed by spin- and angle-resolved photoemission spectroscopy and, additionally, in the case of Au(111) by first principles calculations including spin-orbit coupling \cite{nicolay2001,henk2003,henk2004}.

Large or giant surface state spin splitting of Rashba-Bychkov origin has been found in thin films and alloys, including Pb/Ag(111) \cite{pacile2006,meier2008}, Bi/Ag(111) \cite{ast2007,meier2008}, Ag/Au(111) \cite{cercellier2006}, Bi(001)/Si(111)-7$\times$7 \cite{hirahara2007} and monolayers of Cu, Al, Ag and Au on W(110) and Mo(110) \cite{shikin2008,rybkin2010,shikin2013} and a Ag monolayer on Pt(111) \cite{bendounan2011}.  Large spin-orbit splittings were also observed in the surface states of W(110)-(1$\times$1)H \cite{hochstrasser2002}, Gd \cite{krupin2005}, and W(110) \cite{rybkin2012}.  Spin split surface states on Au(111) \cite{marchenko2012} and Ir(111) \cite{varykhalov2012,marchenko2013} have been shown to survive the adsorption of a graphene overlayer.

Most of these studies focused on shallow states near the Fermi energy.  Early ARPES measurements of the Au(111) and Ag(111) surfaces found evidence for an additional deep intrinsic surface state in the $sd$ projected band gap at a binding energy of 7.2~eV in Ag and 7.8~eV in Au \cite{kevan1987}.  Deep surface states have been observed on Cu surfaces \cite{baldacchini2003,winkelmann2012} and in Ag films, where the progression from discrete quantum well states to surface projected bulk bands could be followed as a function of film thickness \cite{speer2009}.  These studies of Cu and Ag did not report spin-orbit effects.  Recently, giant Rashba effects were observed in deep states of the topological insulator Bi$_2$Te$_2$Se \cite{miyamoto2014}.

In this paper, we investigate spin-orbit coupling in deep $d$-derived Tamm \cite{tamm1932} surface states in 24-layer Au(111) and Ag(111) films.  We characterize a conical intersection (Dirac cone), predicted by our DFT calculations, between a particular pair of surface states.  The Dirac cone we observe lies at a time-reversal invariant $\overline{M}$ point of the surface Brillouin zone (SBZ) in a large surface-projected bulk band gap.  Unlike $\overline{\Gamma}$ point Dirac cones, which are often nearly isotropic, these $\overline{M}$ point cones are strongly anisotropic due to the lower small point group symmetry of the wave vector $\mathbf{k}_{\overline{M}}$.  Our theoretical predictions are in excellent agreement with ARPES measurements which we carry out and present here along the high symmetry planes as well as constant energy slices.  To explain the anisotropy, we derive an effective $\mathbf{k}\cdot\mathbf{p}$ Hamiltonian with distinct longitudinal and transverse Rashba parameters. 

\section{Methods and materials}
\label{sec:methods}

\subsection{Experimental methods}

A single crystal Ag(111) substrate was prepared by repeated cycles of Ar ion sputtering and annealing to 770$^{\circ}$~C.  Au was deposited on the Ag substrate at 150 K and annealed to room temperature to form atomically uniform films estimated to be 24 monolayers thick. The Ag films were grown in a similar way on very thick Au films.  For both systems, low energy electron diffraction showed a sharp 1$\times$1 hexagonal pattern, indicating in particular that the Au(111) film surface remained unreconstructed.

Photoemission spectra were measured at the VUV-Photoemission beamline of  the Elettra synchrotron (Italy) with the samples kept at 150 K, using a Scienta R-4000 electron analyzer and 35 to 80 eV photons. Energy and angular resolution were set at 25 meV and 0.3$^{\circ}$. The data are presented in second derivative to enhance the sensitivity to low intensity features.

\subsection{Computational methods}

Ag(111) and Au(111) thin films were modeled by 24-layer slabs separated by 21~\AA~of vacuum.  This number of layers is sufficient to effectively decouple the top and bottom slab surfaces \cite{forster2007}.  
\begin{figure}[t!]
\includegraphics[width=0.9\columnwidth]{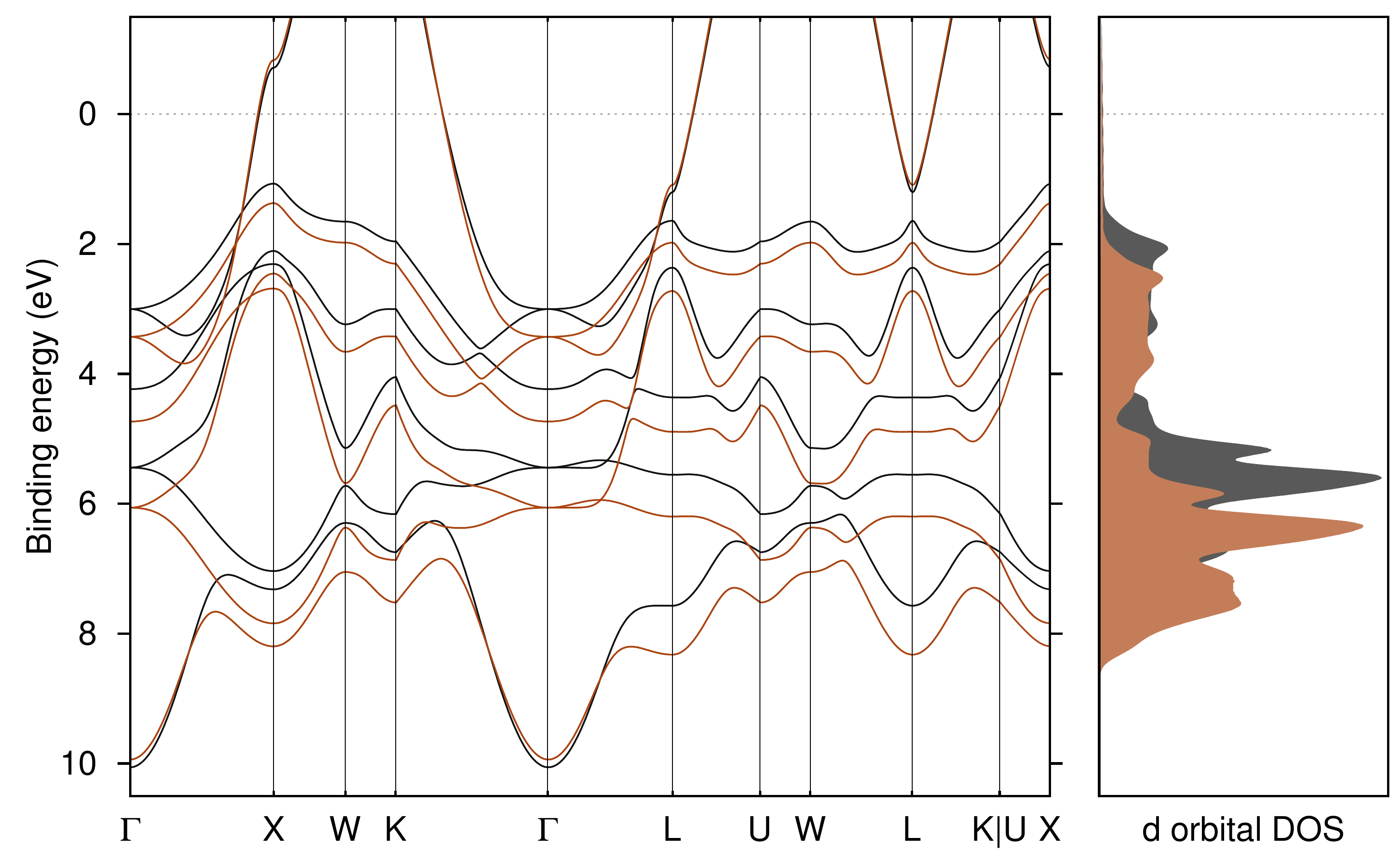}
\caption{DFT band structure of bulk fcc Au in the LDA (black) and LDA+$U$ (orange) approximations.  The value of the Hubbard correction, $U=1.5$~eV, was chosen to shift the center of gravity of the $d$ bands into agreement with experiment.}
\label{fig:Au:bulk}
\end{figure}
Density functional theory calculations in the local density approximation with the parametrization of Perdew and Zunger \cite{perdew1981} were performed with \textsf{QuantumEspresso} \cite{giannozzi2009}, a plane wave pseudopotential electronic structure code.  The lattice constants of the Ag and Au slabs were set according to their bulk experimental lattice constants (4.0782~\AA~for Au and 4.0853~\AA~for Ag), and the interlayer spacing within the slab was held fixed. Brillouin zone integrations were performed on a 12$\times$12$\times$12 $k$-point mesh in bulk Au calculations, a 20$\times$20$\times$20 mesh in bulk Ag calculations, and identical transverse mesh densities in slab calculations, with a smearing width of 0.020~Ry for Au and 0.0075~Ry for Ag.  Plane wave cut-offs were 30 Ry for the wave function and 360 Ry for the charge density. The LDA+$U$ method in the implementation of Ref.~\onlinecite{cococcioni2005} with $U=1.5$~eV was used for Au, while standard LDA was used for Ag.  The value of the Hubbard parameter was chosen to shift the center of gravity of the Au $d$ bands down by 0.55~eV in accordance with the photoemission spectra.  The resulting shift in the band structure is shown in Fig.~\ref{fig:Au:bulk}.  A similar strategy of including Hubbard corrections through a renormalization of the nonlocal pseudopotential coefficients rather than a shift of the atomic orbitals has been applied to one-dimensional gold wires \cite{sclauzero2013} to avoid spurious magnetic solutions due to the excessive proximity of $d$ bands to the Fermi energy, caused in turn by self-interaction errors in standard semilocal approximations.

\section{Dirac cones in deep noble metal surface states}
\label{sec:dirac}

The deep states between 3 and 10~eV binding energy below the Fermi level on noble metal surfaces have mainly $d$ character, and thus large spin-orbit induced band splittings.  In this section, we investigate the effects of spin-orbit coupling on deep noble metal surface states and characterize an unusual Dirac cone at the $\ov{M}$ point of the surface Brillouin zone, explaining the origin of its anisotropic spin splitting.  Our photoemission spectra of Au(111) and Ag(111) films actually reveal an intricate set of surface states and surface resonances, including several of those predicted in Ref.~\onlinecite{mazzarello2008}.

\subsection{ARPES and DFT results for the Dirac cone at the $\ov{M}$ point of Au(111)}
\label{ssec:dirac:au}

\begin{figure}[t!]
\includegraphics[width=0.95\columnwidth]{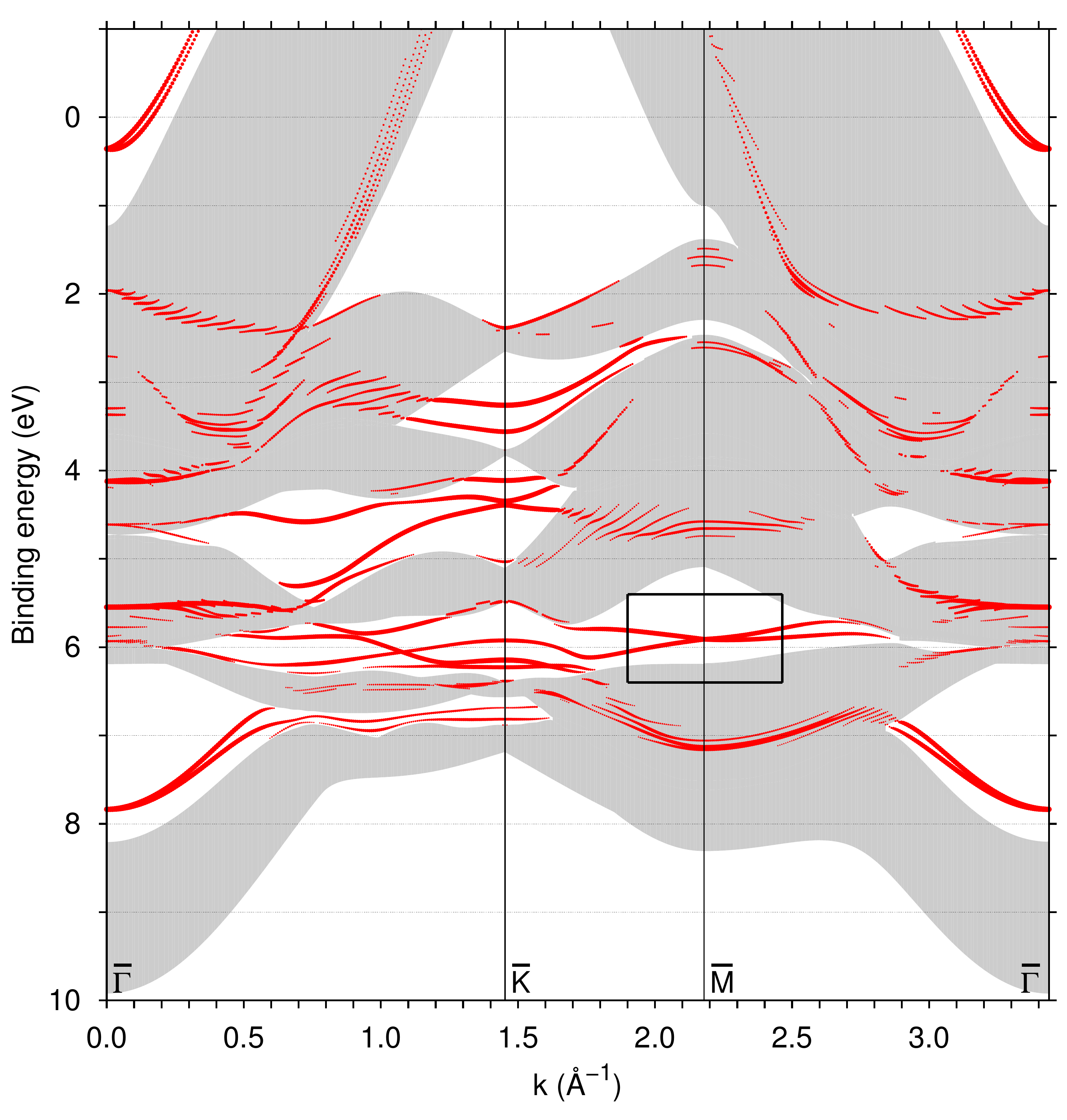}
\caption{Au(111) surface: DFT electronic band structure along the path $\overline{\Gamma KM\Gamma}$ with surface states highlighted in red according to the amplitude of the state on the top two surface layers; shaded areas show the surface-projection of bulk bands. The box surrounds the Dirac cone to be described later.}
\label{fig:Au:GKMG}
\end{figure}

Figure~\ref{fig:Au:GKMG} shows the fully relativistic DFT band structure of a 24-layer Au(111) film for binding energies down to 8.5~eV with the surface states highlighted in red according to their amplitude on the top two surface layers.  The $\ov{M}$ point Dirac cone is located inside the box at 5.9~eV binding energy.

The well-known Rashba spin splitting of the $L$-gap Shockley surface state beginning 0.45~eV below the Fermi energy is visible at the top of the graph. Spin-orbit interactions have a stronger and more varied effect on deep states, splitting the $d$ orbital surface state bands according to their spin polarization by up to 1.5~eV and opening additional band gaps hosting new surface states \cite{mazzarello2008}.

\begin{figure}[t!]
\includegraphics[width=0.85\columnwidth]{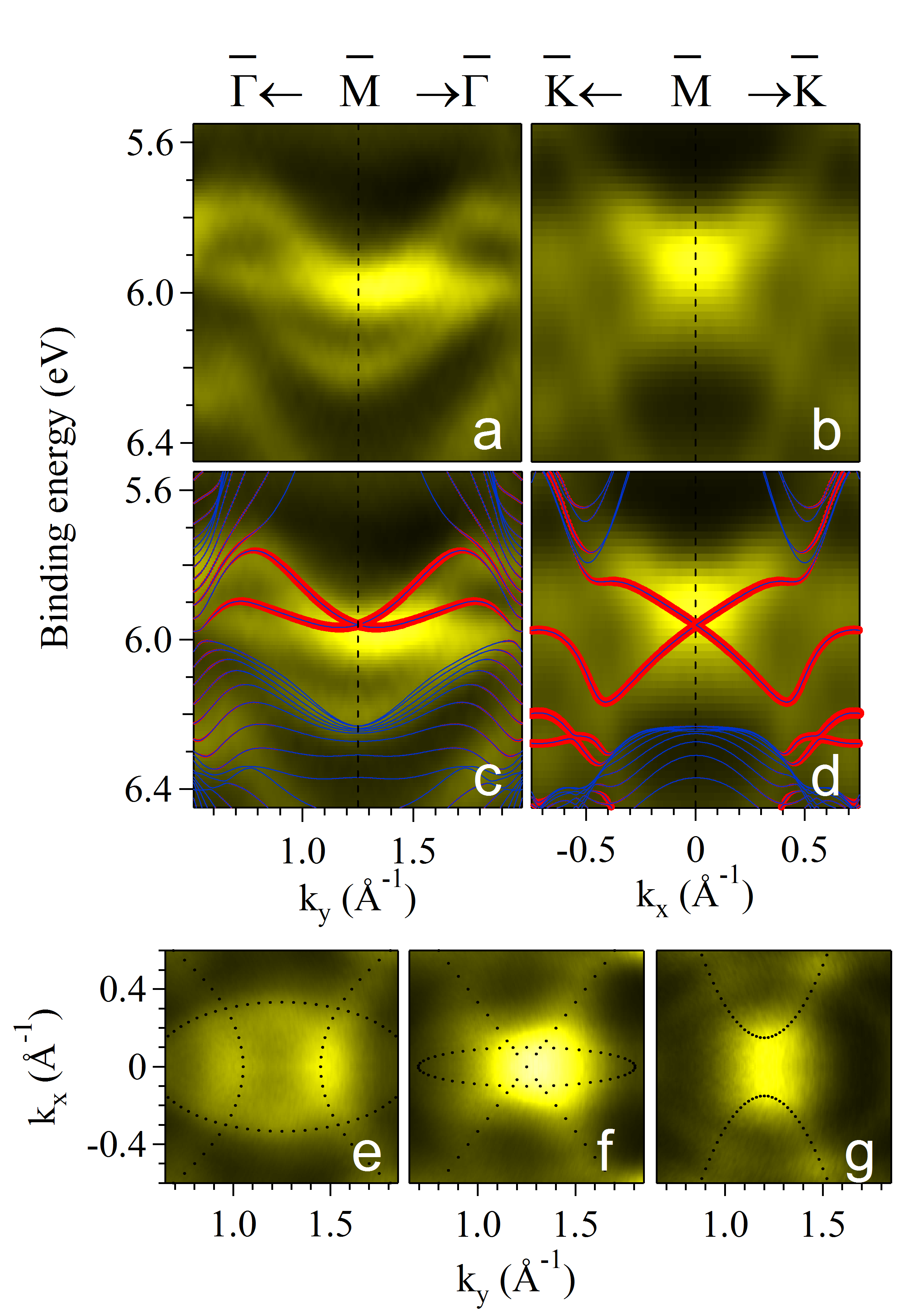}
\caption{Dirac cone in surface states of the Au film/Ag(111):  (a-b) ARPES data; (c-d) DFT calculations (colored lines, red = surface amplitude) with ARPES data in background.  Bands are shown along high symmetry lines (panels a-d) and and as constant energy cuts (bottom panel) at binding energies $E-E_D=$(-0.14, 0.0, +0.1)~eV (left, middle, right).  Dotted lines indicate where the bands intersect the constant energy plane. Spectra were measured at a photon energy of 80~eV.}
\label{fig:Au:dirac}
\end{figure}

One of the nontrivial effects of spin-orbit coupling is the formation of a conical intersection between two surface states at the $\ov{M}$ point in a large spin-orbit-induced gap between 5.1 and 6.2~eV.  Figure \ref{fig:Au:dirac} shows ARPES and DFT spectra of Au near $\ov{M}$ along the $\ov{\Gamma}\ov{M}\ov{\Gamma}$ and $\ov{K}\ov{M}\ov{K}$ high symmetry lines.  The agreement between theory (without adjustable parameters)  and experiment is excellent.  Near the Dirac cone the band splitting is linear in $|\mathbf{k}-\mathbf{k}_{\ov{M}}|$, but the constant of proportionality (Rashba parameter) takes different values, $\alpha_{L}$ and $\alpha_{T}$, in longitudinal ($\ov{M}\rightarrow\ov{\Gamma}$) and transverse ($\ov{M}\rightarrow\ov{K}$) directions.  The linear splitting is most clearly visible in the transverse direction; in the longitudinal direction the splitting is also linear but only in a small region near the conical intersection, since the lower band quickly acquires strong upward dispersion.  The splitting in the longitudinal direction resembles the splitting of the $L$-gap states, being a sum of linear and quadratic terms.  The anisotropy is a consequence of the low symmetry ($C_s$) of the small group of the wave vector $\mathbf{k}_{\ov{M}}$.  General group theory arguments indeed show that  anisotropic Rashba splitting can occur at the time-reversal invariant $\ov{M}$ points of the 2d Brillouin zone of a triangular lattice \cite{oguchi2009}.  Anisotropic Rashba splitting was also predicted at the $\ov{Y}$ point of the Au(110) surface \cite{simon2010}, where the small group is $C_{2v}$.

\begin{figure}[t!]
\begin{tabular}{@{}rl@{}}
\includegraphics[width=0.7\columnwidth]{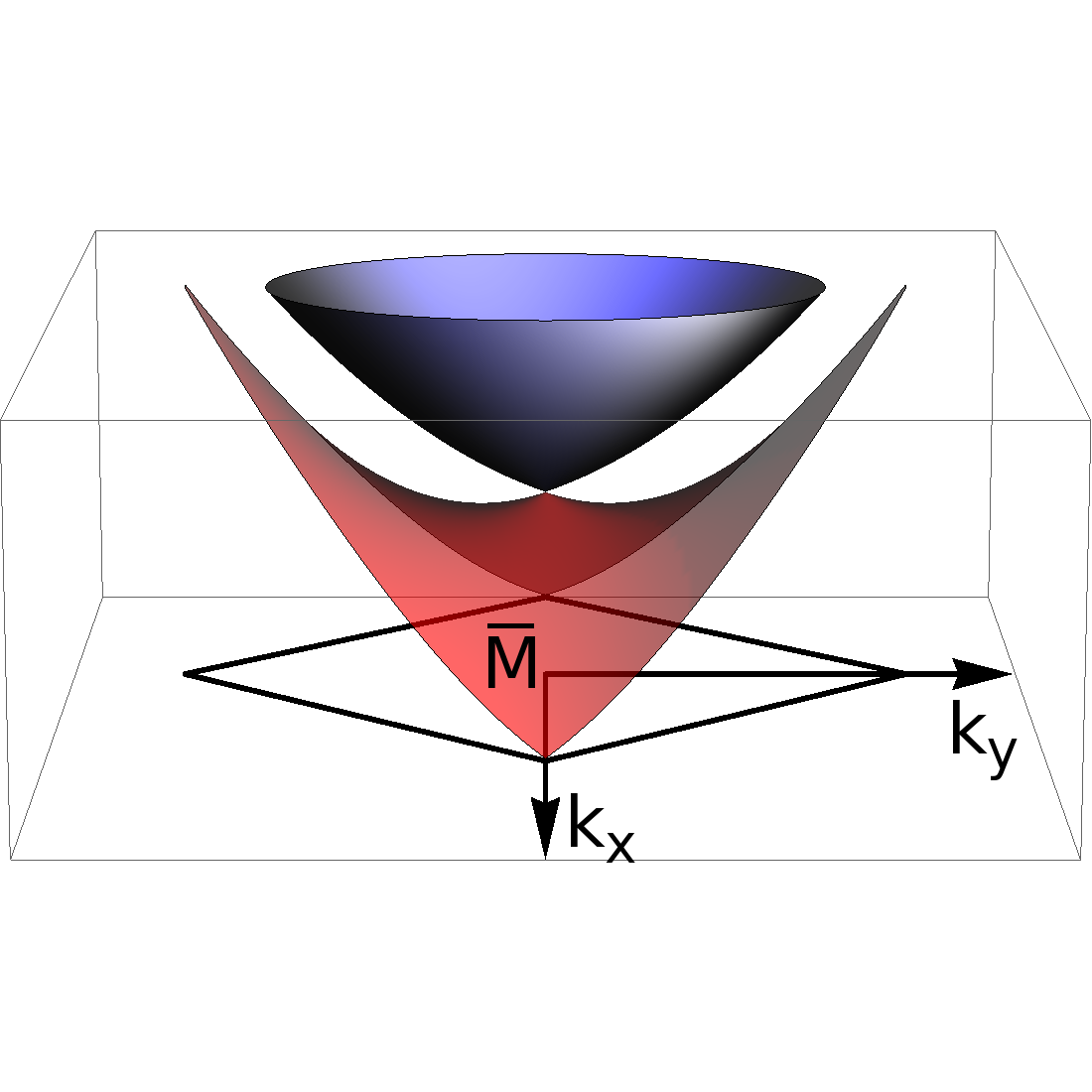} \includegraphics[width=0.25\columnwidth]{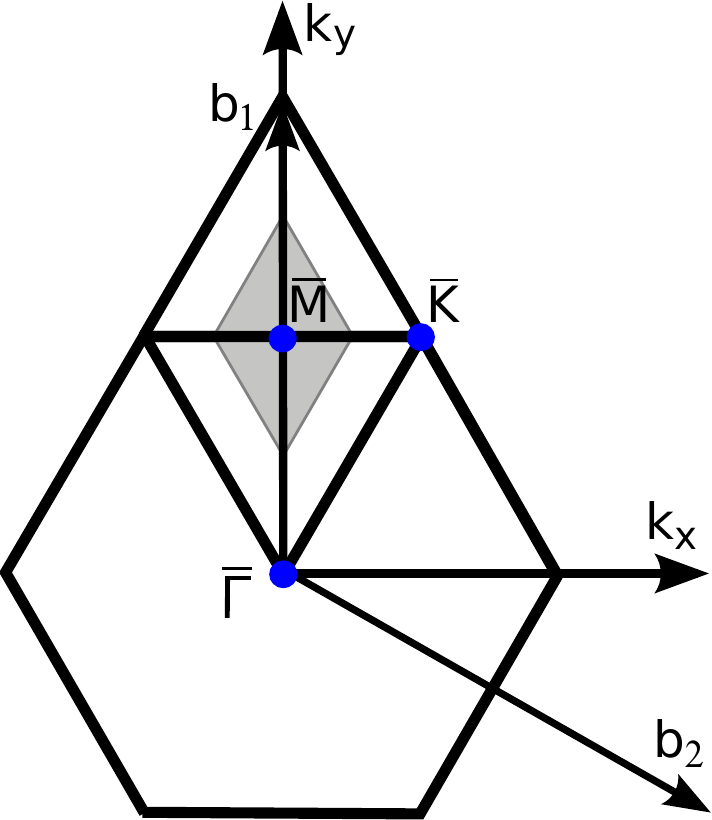}
\end{tabular}
\caption{(left) Warping of the Dirac cone shown schematically; (right) two-dimensional Brillouin zone and choice of reciprocal lattice vectors $\vec{b}_1$ and $\vec{b}_2$.}
\label{fig:warping}
\end{figure}

Away from the $\ov{M}$ point, the lower branch quickly acquires a strong upward curvature.  Accordingly, the Dirac cone displays a strong rhombic warping, shown pictorially in Fig.~\ref{fig:warping}, which mirrors the upward (downward) dispersion of the nearby bulk bands in the longitudinal (transverse) direction.  This warping of the Dirac cone is confirmed by a series of constant energy slices in the lower panels of Fig.~\ref{fig:Au:dirac}, showing the photoemission intensity in a region surrounding $\ov{M}$.  
The circular shape of the upper cone is visible for energies above the conical intersection, i.e.~at lower binding energy than the point of conical intersection.  The enhanced intensity in the longitudinal direction (along $k_x$) indicates that both the upper and the lower branches of the cone are contributing to the signal there, since the lower branch has curved upward by that point.
Likewise, for energies below the Dirac cone, we observe intensity from the lower branch in the transverse direction but not in the longitudinal direction (since the lower branch turns upward before reaching that low in energy).  As seen in Fig.~\ref{fig:Au:GKMG}, the upper branch of the Dirac cone merges with the projected bulk $d$ bands along $\ov{M}\ov{K}$.  The behavior along $\ov{M\Gamma}$ is more difficult to determine since there is a pinching of the upper and lower bulk bands halfway between $\ov{M}$ and $\ov{\Gamma}$ in the DFT calculation.

\subsection{Anisotropy of the Dirac cone at the $\ov{M}$ point}
\label{ssec:anisotropy}

\begin{figure}[t!]
\includegraphics[width=0.9\columnwidth]{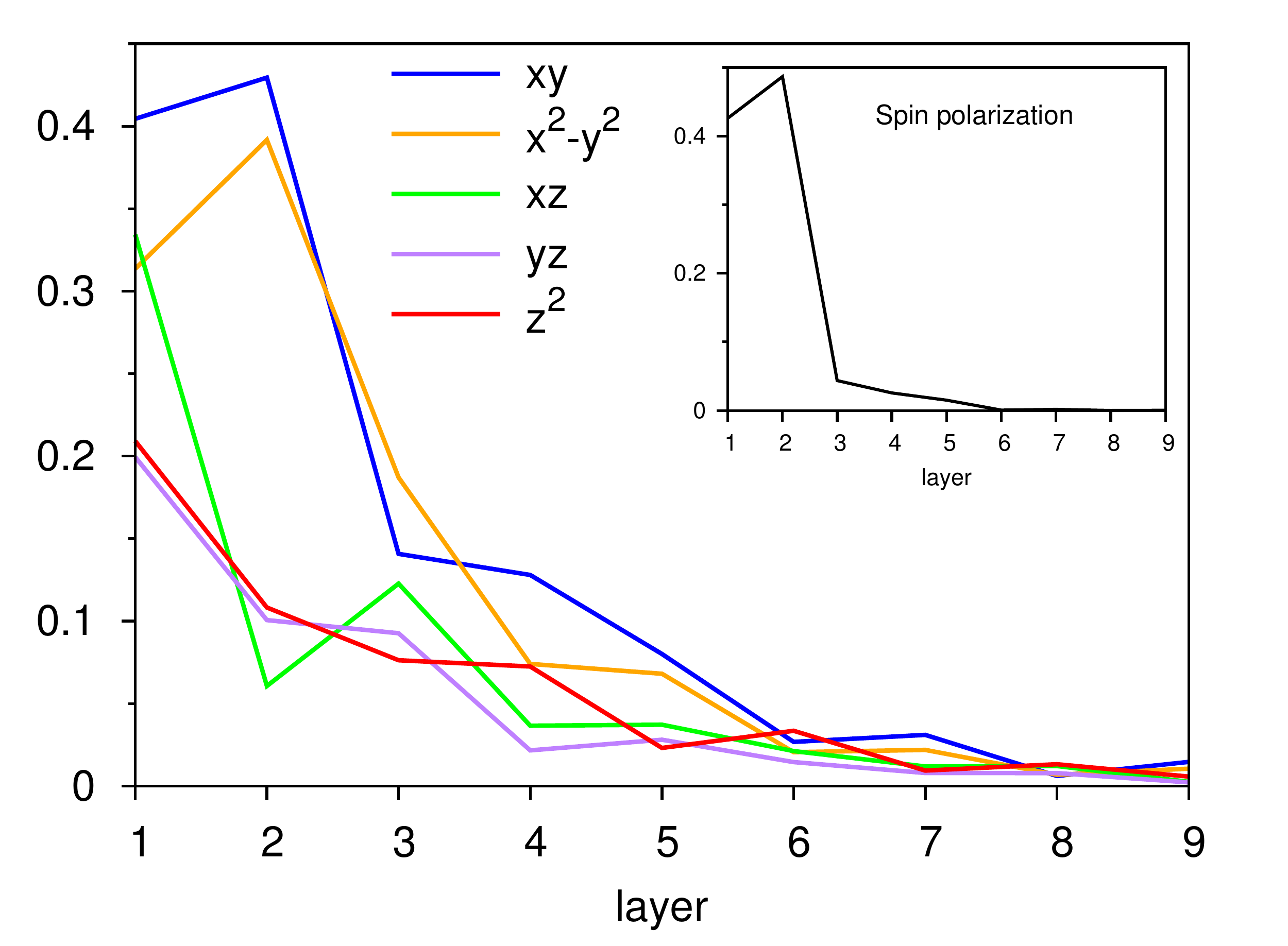}
\caption{Surface state $d$ orbital composition and spin polarization (inset) as a function of layer.}
\label{fig:layer:composition}
\end{figure}

The $\ov{M}$ points of the SBZ belong to a plane of mirror symmetry.  For definiteness, we refer to the $\ov{M}$ point shown in the inset of Fig.~\ref{fig:warping} which belongs to a $yz$ mirror plane.  The Dirac cone surface states are degenerate by time-reversal symmetry at $\ov{M}$ and transform as $\Gamma_3 \oplus \Gamma_4$ irreducible representations of the $C_s$ small double group of the wave vector $\mathbf{k}_{\ov{M}}$.  Near $\ov{M}$, the surface state band dispersion can be described by an effective $\mathbf{k}\cdot\mathbf{p}$ Hamiltonian
\begin{align}
H_{\ov{M}}(\mathbf{k}) = \left( \begin{array}{rr} \alpha_L k_y & \alpha_T k_x \\[0.15cm] \alpha_T k_x & -\alpha_L k_y \end{array} \right) + \frac{\hbar^2 k_x^2}{2m_T^*} + \frac{\hbar^2 k_y^2}{2m_L^*}, \label{eq:kp}
\end{align}
where $k_x$ and $k_y$ refer to transverse and longitudinal displacements from $\ov{M}$ towards $\ov{K}$ and $\ov{M\Gamma}$ respectively.  The general form of the linear Rashba Hamiltonian for a pair of time-reversal degenerate states in $C_s$ symmetry was derived in Ref.~\onlinecite{oguchi2009}.  The longitudinal and transverse Rashba parameters and effective masses $m^*_L$ and $m^*_T$ obtained from fitting our DFT results are reported in Table~\ref{table:params}. 
\begin{table}[th!]
\caption{Rashba parameters and effective masses}
\begin{tabular*}{4.4cm}{l@{\hspace{1.4em}}r@{\hspace{1.5em}}r}
\toprule
  		& Au(111)   		& Ag(111)		\\[0.03cm] \hline
$\alpha_L$ 	& 0.18~eV\AA	& 0.06~eV\AA	\\[0.03cm]
$\alpha_T$	& 0.42~eV\AA	& 0.32~eV\AA	\\[0.05cm]
$m^*_L$	& 150~$m_e$	& 160~$m_e$	\\[0.04cm]
$m^*_T$ 	&-610~$m_e$	& 8300~$m_e$
\end{tabular*}
\label{table:params}
\end{table}
The effective masses are much larger than for the $L$-gap surface states.  Equation~(\ref{eq:kp}) is expressed in terms of a time-reversal pair of basis states $|\psi^{(\Gamma_3)}\rangle = |\psi^{(\Gamma_3)}(\mathbf{k}_{\ov{M}})\rangle$ and $|\psi^{(\Gamma_4)}\rangle = |\psi^{(\Gamma_4)}(\mathbf{k}_{\ov{M}})\rangle$.  These states decay into the bulk and satisfy the Bloch condition with respect to translations by the lattice vectors of the surface.  The longitudinal and transverse Rashba parameters are \cite{oguchi2009}
\begin{equation}
\alpha_{L} = \frac{\hbar}{m} \langle \psi^{(\Gamma_3)} | P_y | \psi^{(\Gamma_3)} \rangle; \qquad \alpha_{T} = \frac{\hbar}{m} \langle \psi^{(\Gamma_3)} | P_x | \psi^{(\Gamma_4)} \rangle {,} \label{eq:alpha}
\end{equation}
where $\mathbf{P} = \mathbf{p} + \frac{\hbar}{4mc^2} \sigma\times\nabla V$.  It is not simple to obtain explicit expressions for the $|\psi^{(\Gamma_3)}\rangle$ and $|\psi^{(\Gamma_4)}\rangle$ basis states, since involve all five $d$ orbitals and extend over multiple layers.  They are made up predominantly of planar $d_{x^2-y^2}/d_{xy}$ orbitals but also have significant admixture of out-of-plane $d_{xz}/d_{yz}$ and $d_{z^2}$ orbitals.  The orbital composition of the lower surface state branch as a function of layer is shown in Fig.~\ref{fig:layer:composition}.
The surface states are more strongly localized near the surface than the $L$-gap surface states.

The inset of Fig.~\ref{fig:layer:composition} shows the fractional contribution of each layer to the total spin polarization $\vec{S}_l\cdot\vec{S}/|\vec{S}|^2$, where $\vec{S}_l$ is the spin polarization of layer $l$ and $\vec{S}$ is the total spin polarization.  Approximately 90 percent of the spin polarization is contributed by the first two layers.  The details of spin polarization will be discussed below.

Similar anisotropic Rashba spin splitting has been predicted at the $\ov{Y}$ point of Au(110) and described through a two-stage perturbative calculation -- first a $\mathbf{k}\cdot\mathbf{p}$ perturbation with respect to $|\mathbf{k}-\mathbf{k}_{\ov{Y}}|$ and subsequently a first order perturbation with respect to the spin-orbit coupling \cite{simon2010}.  This approach yielded expressions for the anisotropic Rashba parameters, but it is not straightforward to apply in the present case of the deep Dirac cone surface states on Au(111) because these states \textit{only} exist when the spin-orbit interaction is nonzero, since the latter is responsible for opening the band gap in which the surface states reside \cite{mazzarello2008}.  The gap is zero before spin-orbit interactions are turned on because two of the bulk bands which project to $\ov{M}$ are symmetry degenerate at the $L$ point of the fcc Brillouin zone, both belonging to the $L_3$ irreducible representation of the $D_{3d}$ small group of $\mathbf{k}_L$.  When spin-orbit interactions are turned on, these bulk bands split into $\Gamma_4$ and $\Gamma_{5,6}$ irreducible representations, i.e.~$L_3 \otimes D_{1/2} = \Gamma_4 + \Gamma_{5,6}$, and the surface states split off from these bands.  To apply $\mathbf{k}\cdot\mathbf{p}$ perturbation theory before opening this gap would require performing degenerate perturbation theory with respect to a highly degenerate reference state, since the density of states has a van Hove singularity at $\ov{M}$.  Therefore, we have performed our calculations in the opposite order, first taking into account spin-orbit interactions and subsequently performing $\mathbf{k}\cdot\mathbf{p}$ perturbation theory with respect to $|\mathbf{k}-\mathbf{k}_{\ov{M}}|$.  The first step leads to the $|\psi^{(\Gamma_3)} \rangle$ and $| \psi^{(\Gamma_4)} \rangle$ basis states in Eq.~(\ref{eq:alpha}).

\begin{figure}[t!]
\includegraphics[width=0.40\columnwidth]{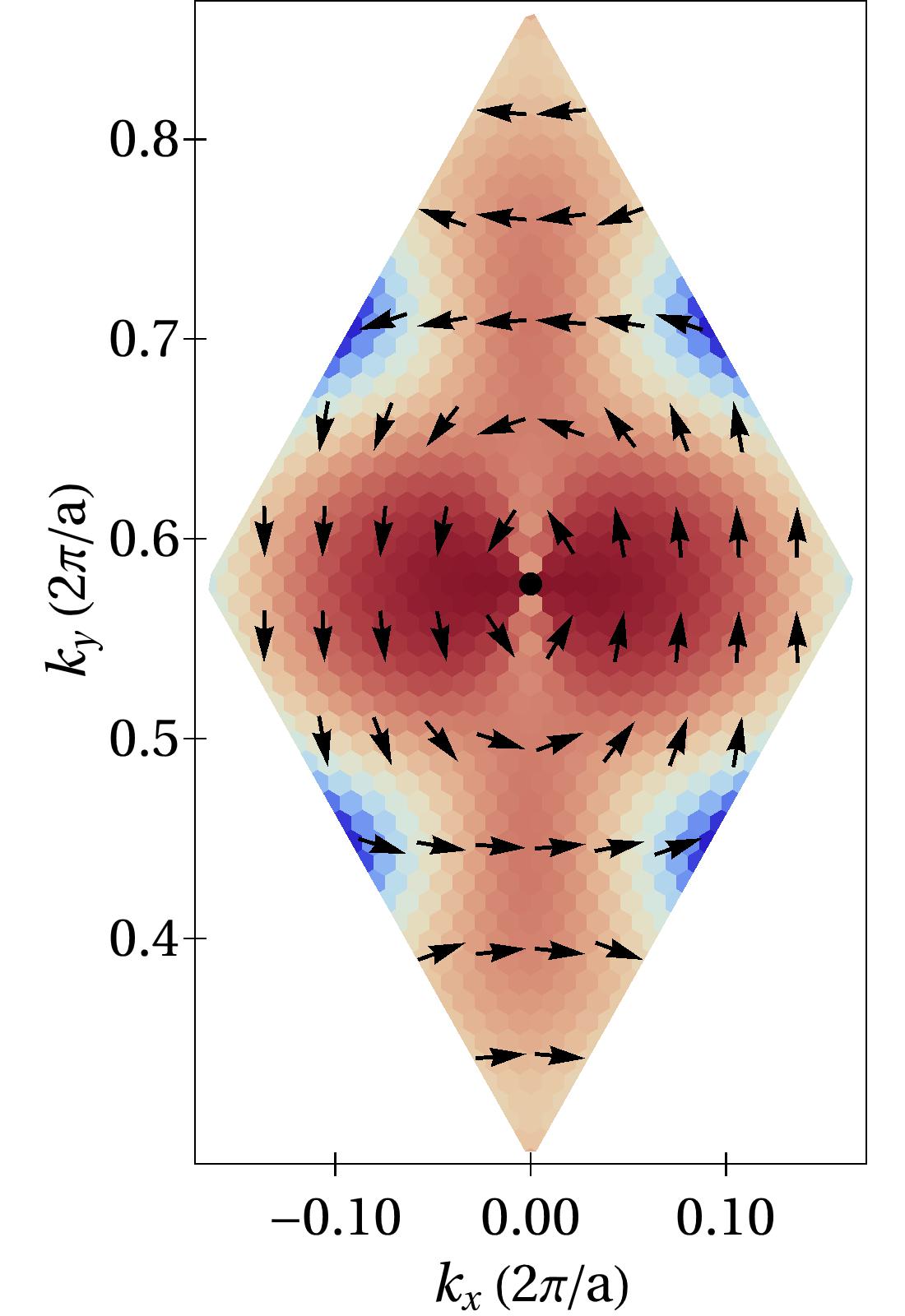} \hspace{0.1cm} \includegraphics[width=0.40\columnwidth]{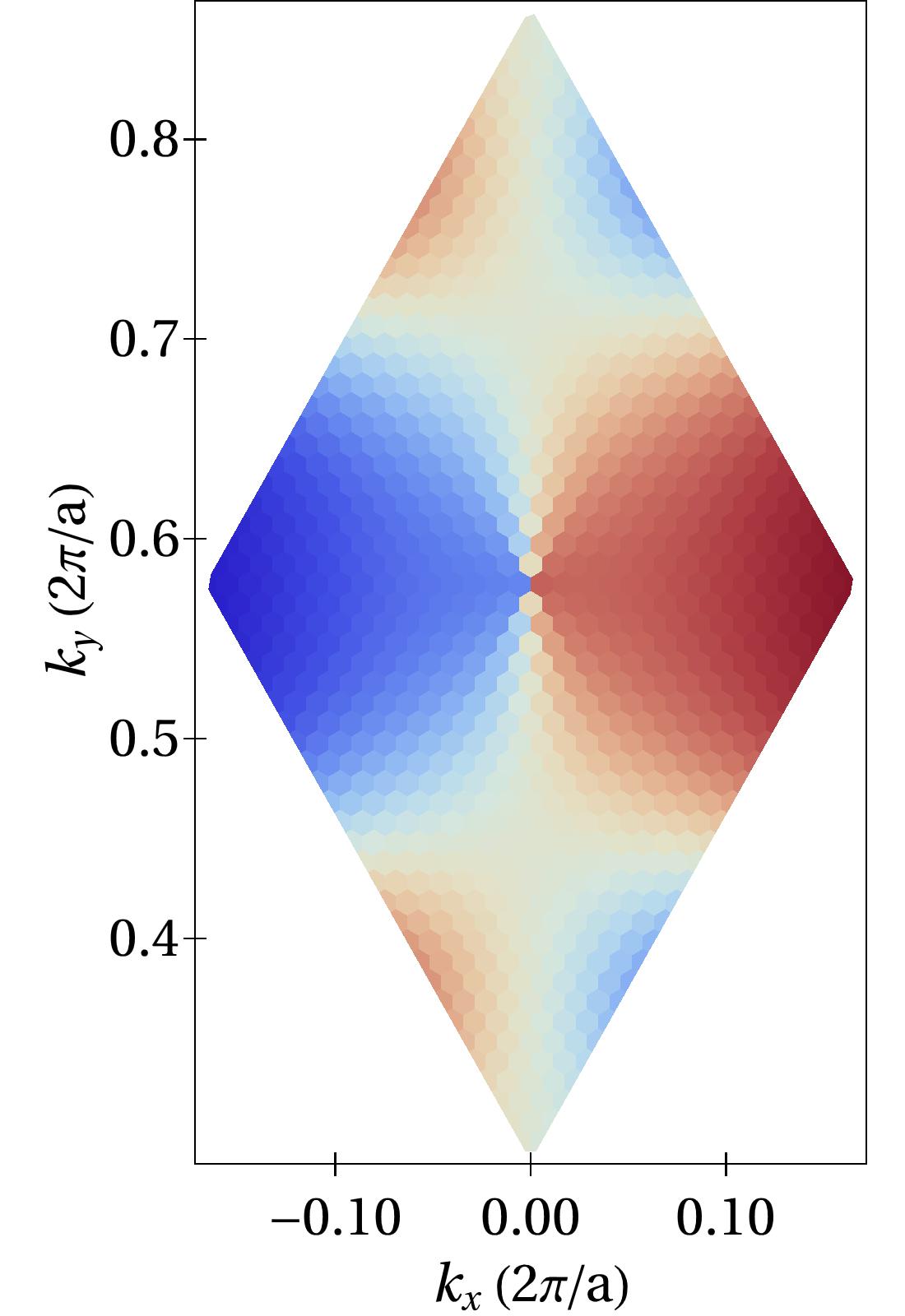} \\
\includegraphics[width=0.40\columnwidth]{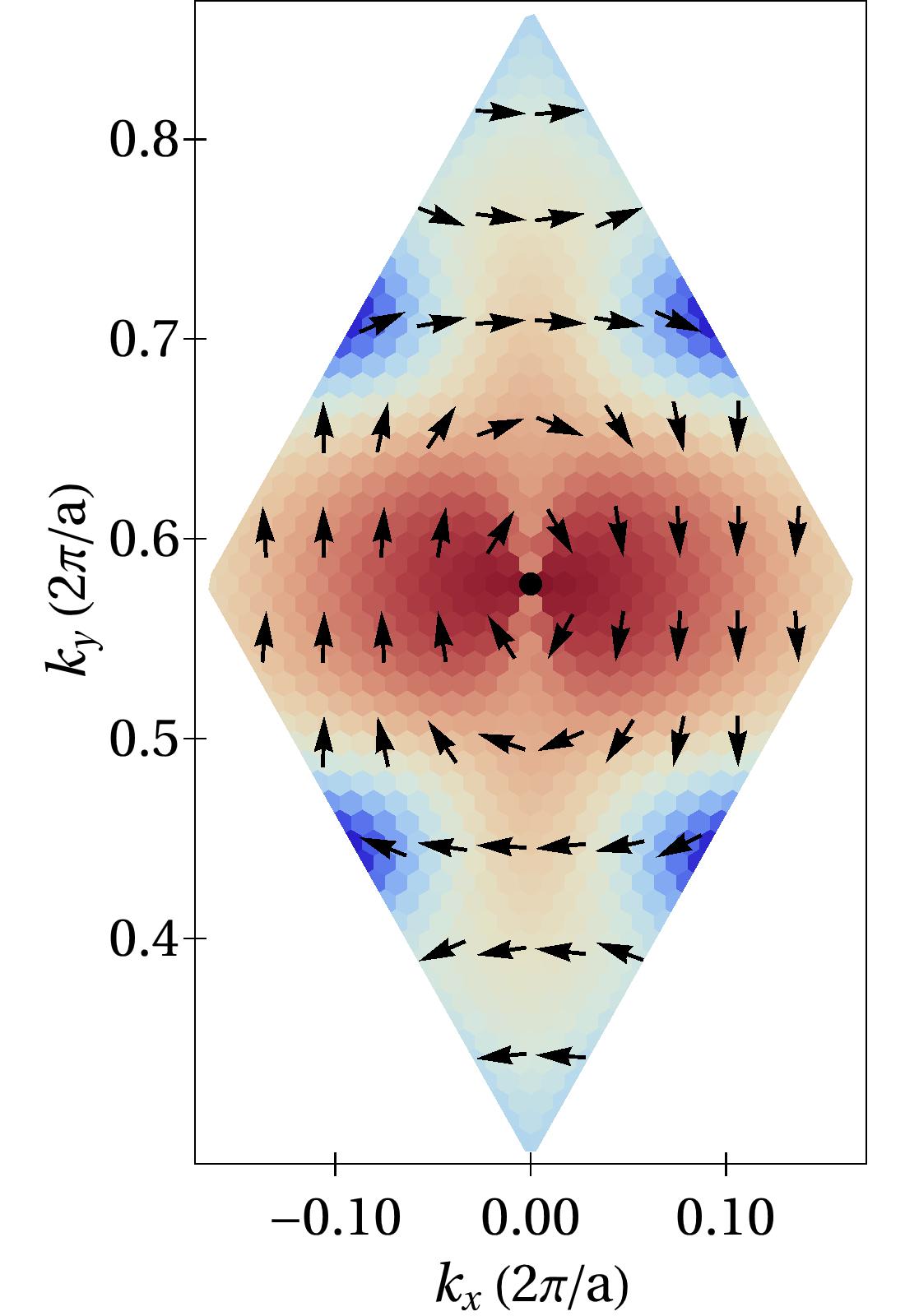} \hspace{0.1cm}\includegraphics[width=0.40\columnwidth]{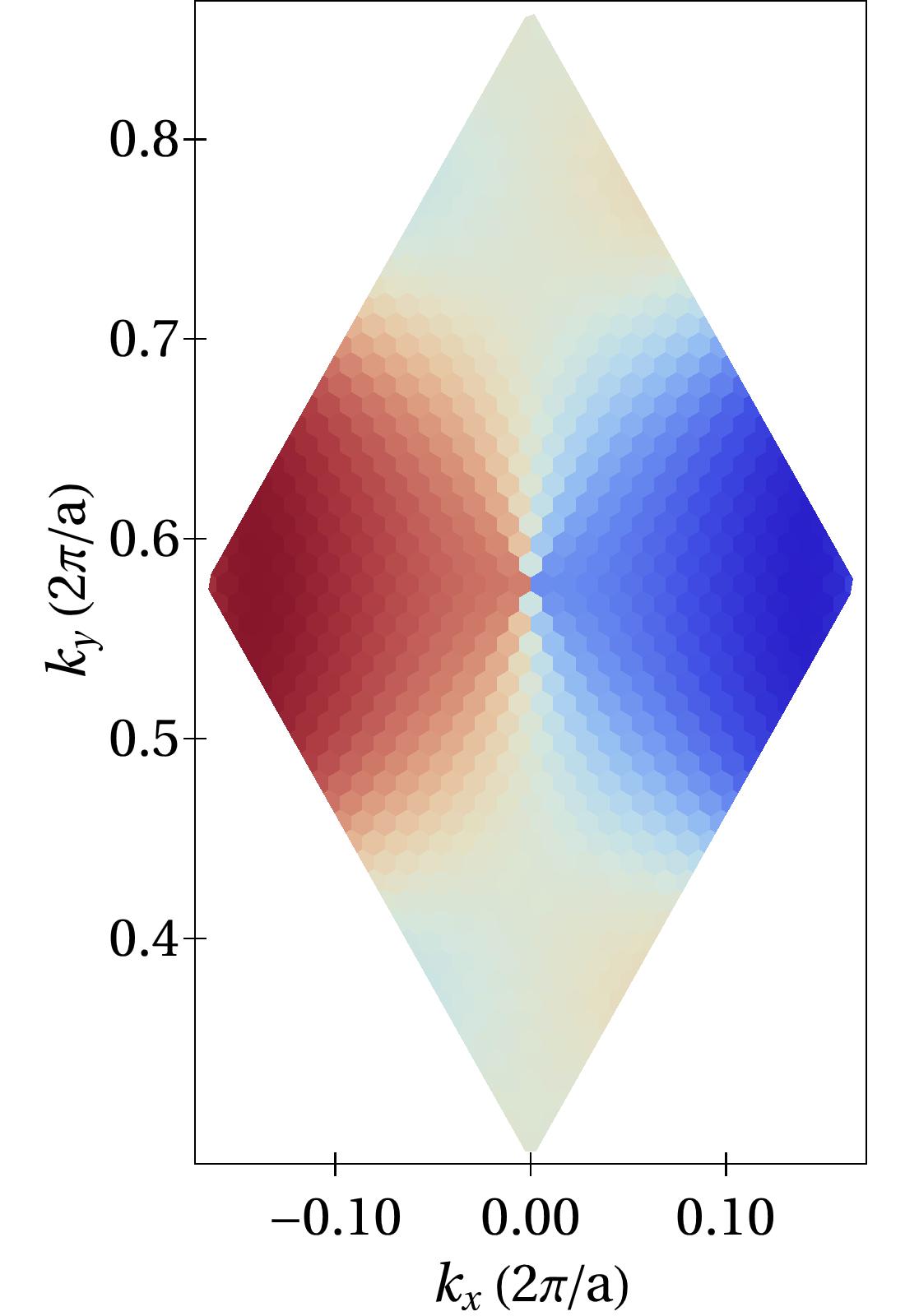} 
\caption{Calculated spin polarization of the upper branch (upper panels) and lower branch (lower panels) of the Dirac cone surface states on an ideal Au(111) surface.  The left panels plot the in-plane spin polarization $(S_x,S_y)$ with the background color showing its modulus (red = maximal, blue = minimal).  The right panels show $S_z$ as a color density plot (red = positive, blue = negative); $S_z/\sqrt{S_x^2+S_y^2}$ ranges between $(-0.75,+0.75)$ for the upper branch and $(-0.85,+0.85)$ for the lower branch.  The plotted region of the surface Brillouin zone is shown in red in the inset.}
\label{fig:Au:spin:polarization}
\end{figure}

Although the ``low energy'' Hamiltonian in Eq.~(\ref{eq:kp}) is adequate for describing the surface states very close to the Dirac cone, it is not valid throughout the SBZ.  The leading corrections to Eq.~(\ref{eq:kp}) are cubic terms $k_x^{\alpha} k_y^{\beta}$ which further contribute to the warping of the Dirac cone.  However, any two-level perturbative approximation for the Dirac cone surface states will break down when they interact with other sets of surface states.  Indeed, the DFT results in Fig.~\ref{fig:Au:GKMG} reveal a complex series of avoided crossings and nonmonotonic spin splitting of the surface state bands.  For instance, multiple sets of surface states interact between $\ov{K}$ and $\ov{M}$.

\begin{figure}[t!]
\includegraphics[width=0.45\columnwidth]{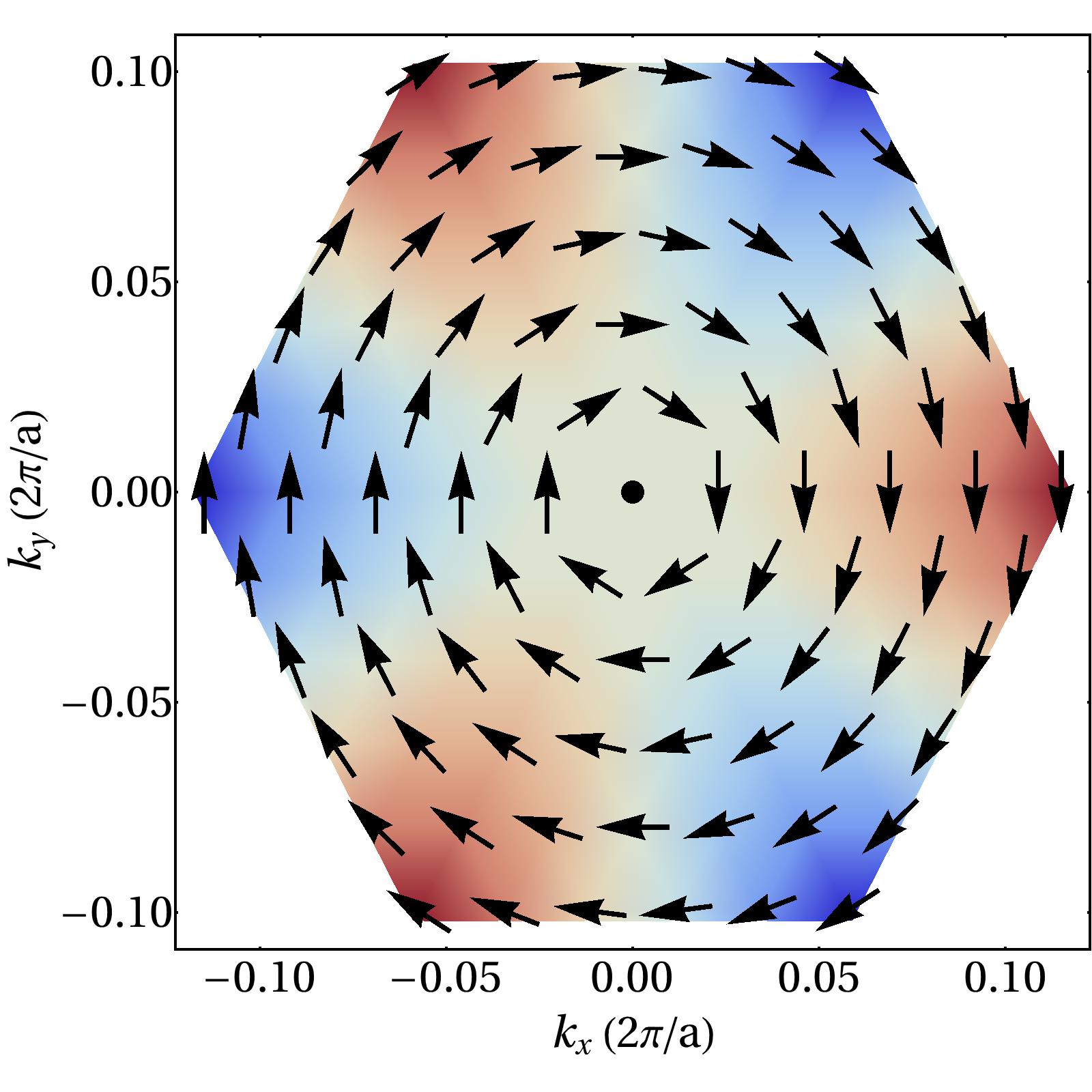} \hspace{0.1cm}\includegraphics[width=0.45\columnwidth]{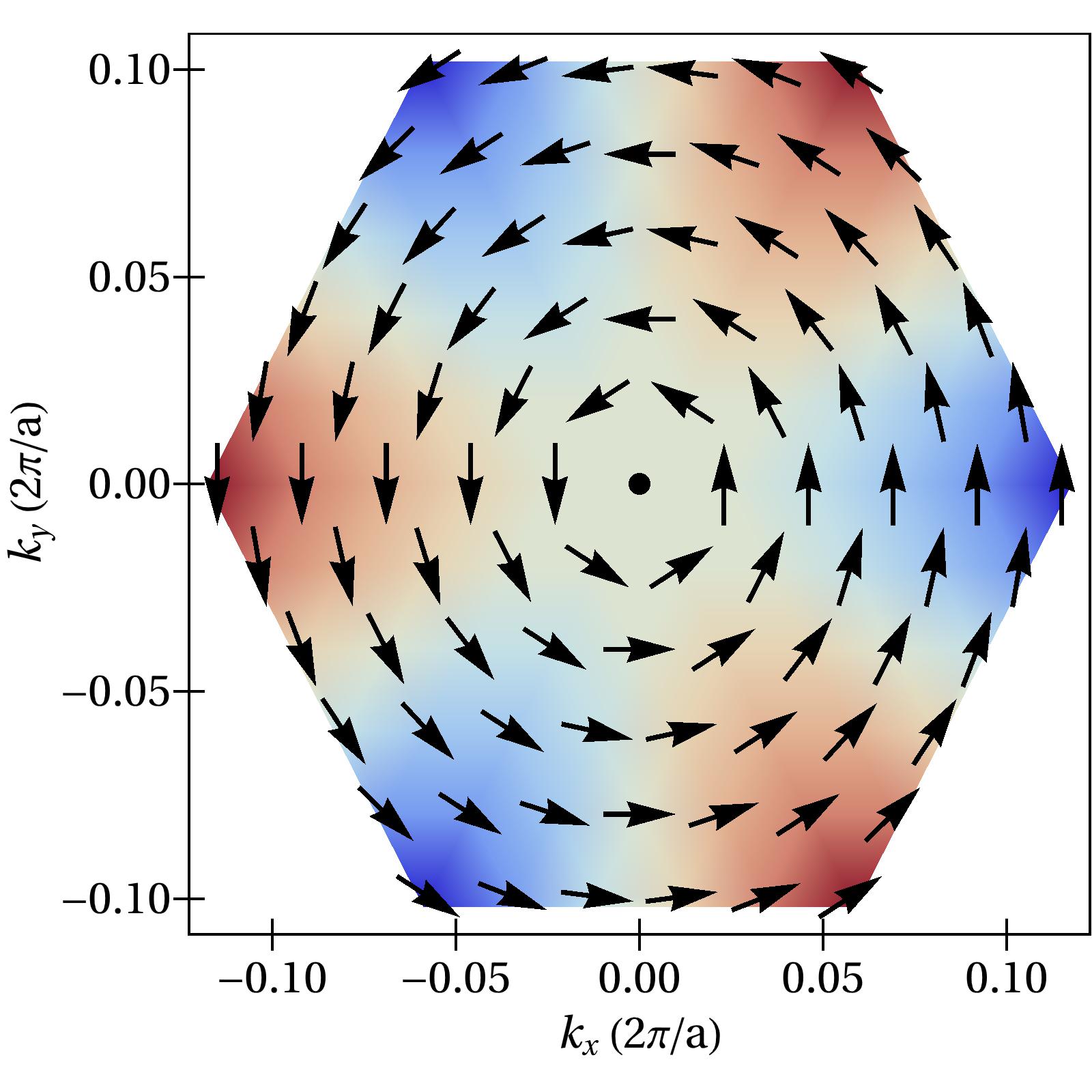}
\caption{Calculated spin polarization of the inner branch (left panel) and outer branch (right panel) of the Dirac cone of the $L$-gap states at $\Gamma$ on the Ag(111) surface.  The $z$ component of spin is shown by the background color (red = positive, blue = negative).}
\label{fig:Ag:$L$-gap}
\end{figure}

The Dirac cone surface states have chiral spin polarization, meaning the spin expectation value $\langle \psi | \vec{S} | \psi \rangle$ is approximately perpendicular to the wavevector $\mathbf{k}-\mathbf{k}_{\ov{M}}$, with opposite handedness for the upper and lower branches, exactly like the Dirac cones of topological surface states and the $L$-gap states of Au.  The in-plane $(S_x,S_y)$ and out-of-plane $S_z$ components of the calculated spin polarization of the upper and lower branches of the surface states are shown in Fig.~\ref{fig:Au:spin:polarization} in a region surrounding the $\ov{M}$ point.  From these plots, it is apparent that the symmetry of the Dirac cone is lower than the approximately hexagonal symmetry of the $L$-gap surface states at $\ov{\Gamma}$.  This is clearly visible by comparing the out-of-plane spin polarization of the $\ov{M}$ point Dirac cone with that of the $\ov{\Gamma}$ point cone shown in Fig.~\ref{fig:Ag:$L$-gap}.  There are significant differences in the spin polarization pattern due to the symmetry lowering from $C_{3v}$ at the $\ov{\Gamma}$ point to $C_s$ at $\ov{M}$.  Additional structure in the spin texture is seen farther away from $\ov{M}$.  The predicted chiral spin polarization should be observable in spin-resolved ARPES measurements.  To further confirm the presence of a conical intersection, we have evaluated the Berry phase for the upper and lower branches of the surface states on a path encircling the conical intersection \cite{berry1984}.  For both branches, we obtained the expected value $\pi\mod 2\pi$ for a two state system, since the states can be chosen to be real in the presence of time reversal symmetry.

Since the $d$ orbital surface states of Au(111) are Tamm surface states, they are generally strongly  localized near the surface, and it is possible to model their band dispersion, spin splitting and spin texture throughout the SBZ by an effective tight binding model on a 2d triangular lattice where spin-orbit interactions are described through spin-flip hopping amplitudes.  Such a model must generally contain non-Hermitian and energy dependent terms (self-energy terms) $PHQ(E-QHQ)^{-1}QHP$ to describe how the surface states transform into surface resonances whenever they merge with bulk bands; $P=\sum_{a} |\psi_a\rangle\langle\psi_a|$ is the projector onto orbitals of the surface layer  and $Q=1-P$.  The minimal spin-orbit interaction term producing spin splitting and Rashba-type chiral spin polarization is
\begin{align}
H_{dd} = \sum_{\langle ij \rangle} i \lambda_{ab} c_{ia\mu}^{\dag} (\vec{\sigma}_{\mu\nu} \times \vec{\mathbf{d}}_{ij})_z c_{jb\nu} {,} \label{eq:Rashba:multiband}
\end{align}
where $\vec{\sigma}$ is the vector of Pauli matrices acting in spin space, the directed bond $\vec{d}_{ij} = \vec{r}_i-\vec{r}_j$ between lattice sites $i$ and $j$, and the $ab$ indices are summed over the manifold of $d$ orbitals; the Einstein summation convention is used for all tensors.  In the case of the deep surface states of Au(111), all $d$ orbitals are relevant.

The spin-orbit interaction term in Eq.~(\ref{eq:Rashba:multiband}) is a multiband generalization of the spin-orbit interactions in the Kane-Mele model \cite{kane2005} with $\lambda_{ab}$ acting as a tensor of generalized Rashba coupling constants; under a global transformation of the local coordinate frame, $\lambda_{ab}$ transforms as $\lambda_{ab} \rightarrow D^{*}_{a\mu} \lambda_{\mu\nu} D_{\nu b}$, where $D_{\mu\nu}=D^{(l=2)}_{\mu\nu}(\alpha,\beta,\gamma)$ is the Wigner $D$-matrix for Euler angles $\alpha$, $\beta$ and $\gamma$.  Initial calculations show that when $H_{dd}$ is added to a 2d tight binding model based on \text{ab initio} parameters, it can describe the independent Rashba-type spin splitting of each individual surface state $d$ band (planar, out-of-plane, etc).  With the addition of next-nearest neighbor intrinsic spin-orbit coupling terms, the tight binding model is further able to describe the surface state spin splitting and spin polarization throughout the SBZ, not only its characteristic linear Rashba spin splitting at a Dirac cone.  Diagonal elements such as $\lambda_{xz,xz}$ and $\lambda_{xy,xy}$ can be decomposed to $dd\sigma$, $dd\pi$ and $dd\delta$ spin-flip hopping contributions.  Off-diagonal elements such as $\lambda_{xz,yz}$ and $\lambda_{x^2-y^2,xy}$ are generally complex and represent genuine multiorbital spin-orbit interactions necessary to describe the surface state dispersion and spin texture throughout the SBZ.  A tight binding model for $d$ orbitals with multiband spin-orbit interactions given by $H_{dd}$ provides a concrete realization of a non-Abelian gauge theory.

The mechanism responsible for inducing spin-orbit interactions in the $d$-orbital Tamm surface states of Ag and Au is different than the one for the free electron-like $L$-gap surface states.  To understand the mechanism, tight binding models similar to those constructed for spin-orbit effects in the $L$-gap surface states of Au \cite{petersen2000} and the $\pi$ bands of graphene \cite{min2006,konschuh2010,konschuh2012,ast2012}, where spin-flip hopping is generated through a combination of electric-field induced hybridization, introduced through matrix elements like  $z_{sp}=\langle s | z | p_z\rangle$, and on-site $\mathbf{L}\cdot\mathbf{S}$ coupling, are more relevant.  However, in the $d$ orbital surface states of Ag and Au, the hybridization with $s$ and $p$ orbitals due to the intrinsic electric field at the surface plays only a minor role in generating the spin splitting, and the relevant spin-flip hopping amplitudes in effective 2d lattice models are instead induced by the downfolding of interlayer hopping and on-site $\mathbf{L}\cdot\mathbf{S}$ coupling, as will be described elsewhere.

\subsection{ARPES and DFT results for the Dirac cone at the $\ov{M}$ point of Ag(111)}
\label{ssec:dirac:ag}

\begin{figure}[t!]
\includegraphics[width=0.95\columnwidth]{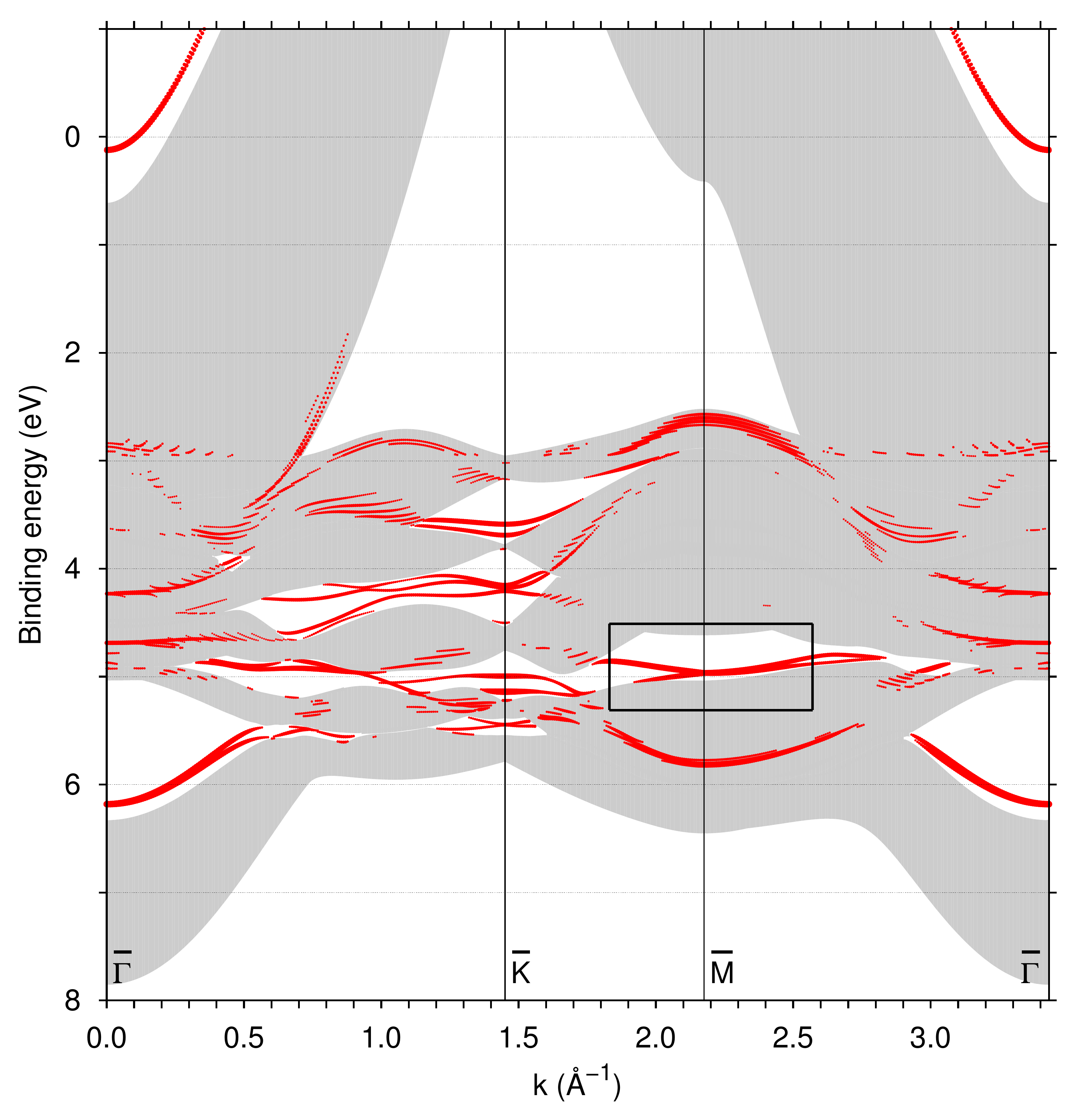}
\caption{Au(111) surface: DFT electronic band structure along the path $\overline{\Gamma KM\Gamma}$ with surface states highlighted in red according to the amplitude of the state on the top two surface layers; shaded areas show the surface-projection of bulk bands.}
\label{fig:Ag:GKMG}
\end{figure}

The above results were for the Au(111) film surface.  Turning our attention now to the lighter noble metal Ag(111) surface, we find that the same Dirac cone occurs in the corresponding surface states despite the weaker spin-orbit interaction.  The fully relativistic DFT band structure for Ag(111) is plotted in Fig.~\ref{fig:Ag:GKMG}, where the Dirac cone is visible in the $\ov{M}$ point gap between 4.6 and 5.0~eV binding energy.  Most of the surface states seen in Au are also present in Ag.  A few of these states have been observed in Ref.~\onlinecite{speer2009}.  The spin-orbit induced projected band gaps in Ag are reduced by a factor between two and three with respect to Au.

\begin{figure}[t!]
\includegraphics[width=0.85\columnwidth]{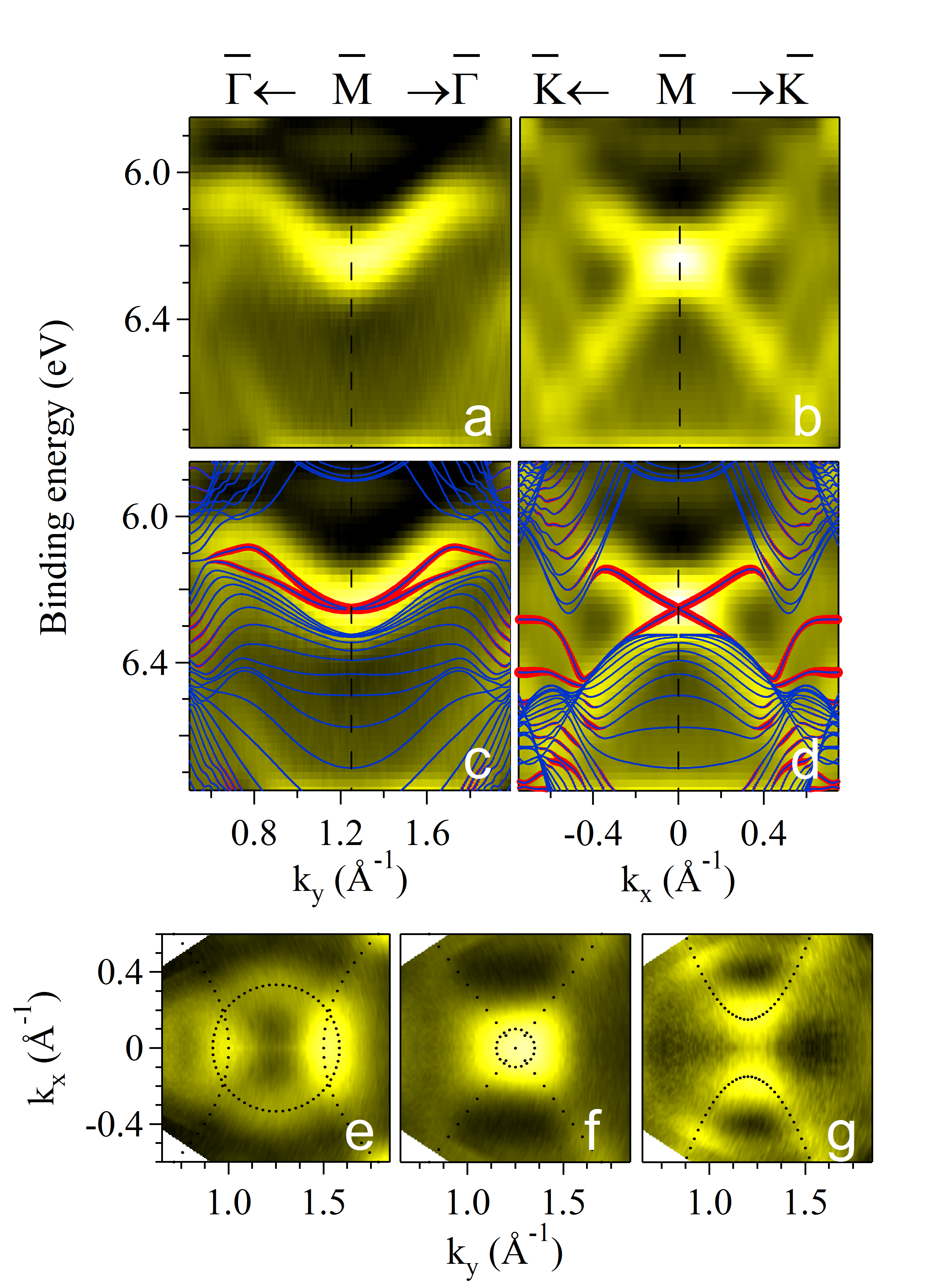}
\caption{Dirac cone in surface states of a Ag film/Au(111):  ARPES (background) and DFT (colored lines, red = surface amplitude) bands are shown along $\ov{\Gamma}\ov{M}\ov{\Gamma}$ (left panels) and $\ov{K}\ov{M}\ov{K}$ (middle panels) high symmetry lines and along constant energy cuts (bottom panels).  Dotted lines indicate where the bands intersect the constant energy plane.  Spectra were measured at a photon energy of 80~eV and DFT bands were shifted down rigidly by 1.29~eV.}
\label{fig:Ag:dirac}
\end{figure}

The Dirac cone of the deep surface states of Ag is shown at higher resolution in Fig.~\ref{fig:Ag:dirac}.  Since the calculated $d$ band energies are again too high with respect to experiment, they have been rigidly shifted down by 1.29~eV in order to obtain the best agreement with experiment; this energy shift does not affect the band dispersion in any way.  The splitting of the surface bands is clearly visible in the transverse $\ov{M}\rightarrow\ov{K}$ direction but not in the longitudinal $\ov{M}\rightarrow\ov{K}$ direction.  Surprisingly, the transverse Rashba parameter $\alpha_{T} = 0.32$~eV~\AA~is almost as large as it was in Au, and therefore the splitting along $\ov{M}\ov{K}$ is more easily resolved.  Its smallness might indicate that induced $dd\pi$ type spin-flip hopping amplitudes are smaller in Ag than Au.  The transverse effective masses is very large.

\begin{figure}[t!]
\includegraphics[width=0.40\columnwidth]{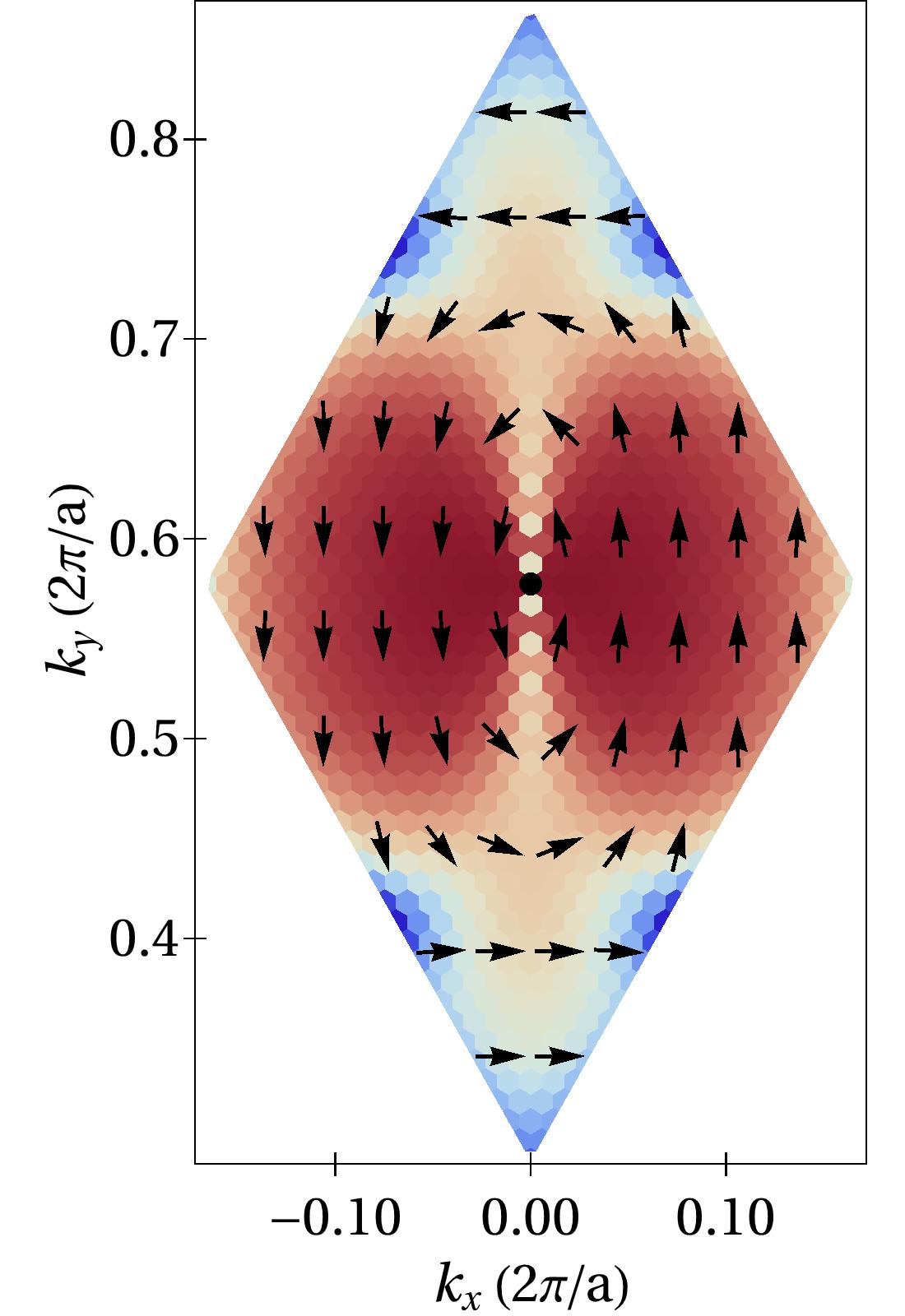} \hspace{0.1cm} \includegraphics[width=0.40\columnwidth]{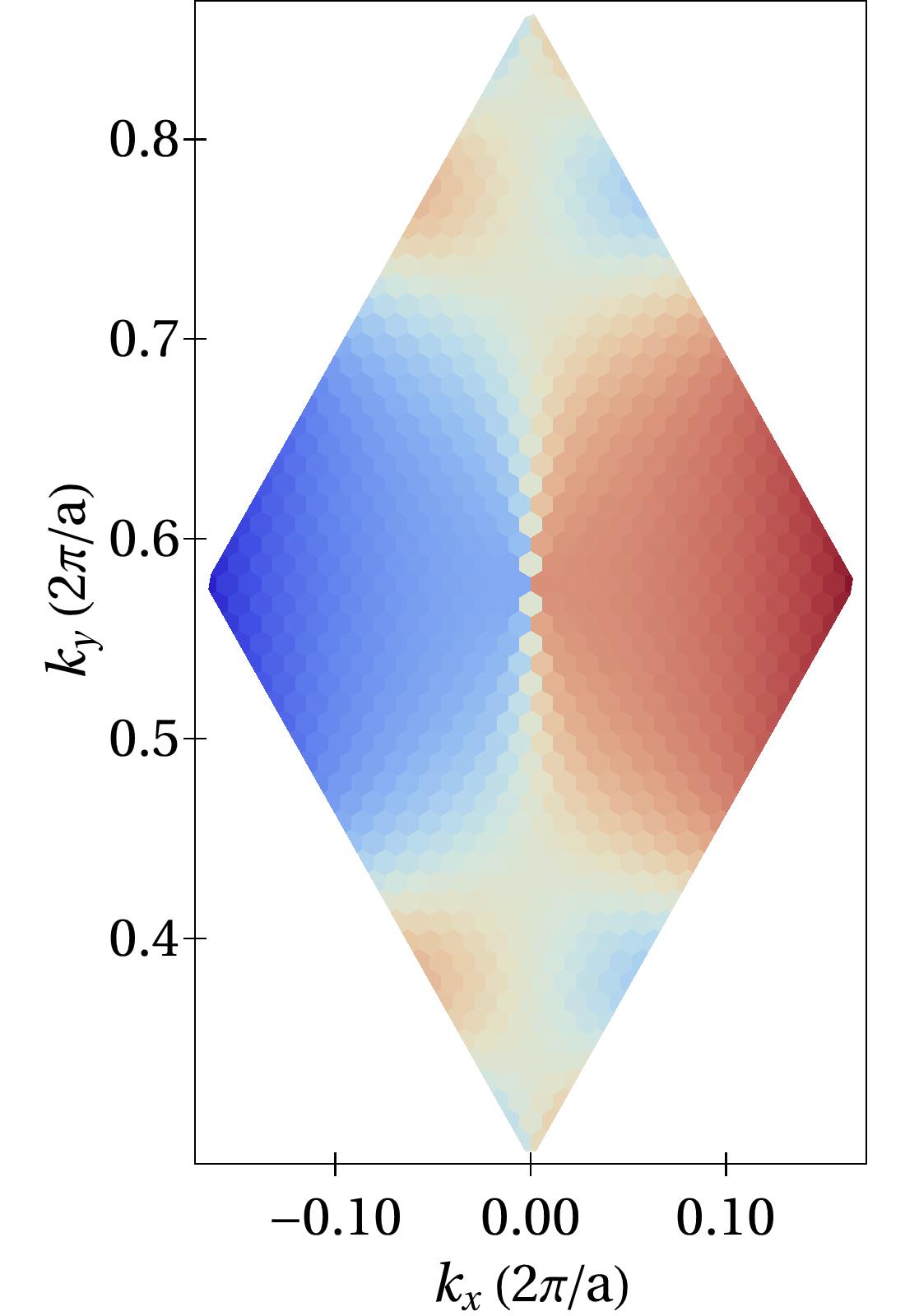} \\ 
\includegraphics[width=0.40\columnwidth]{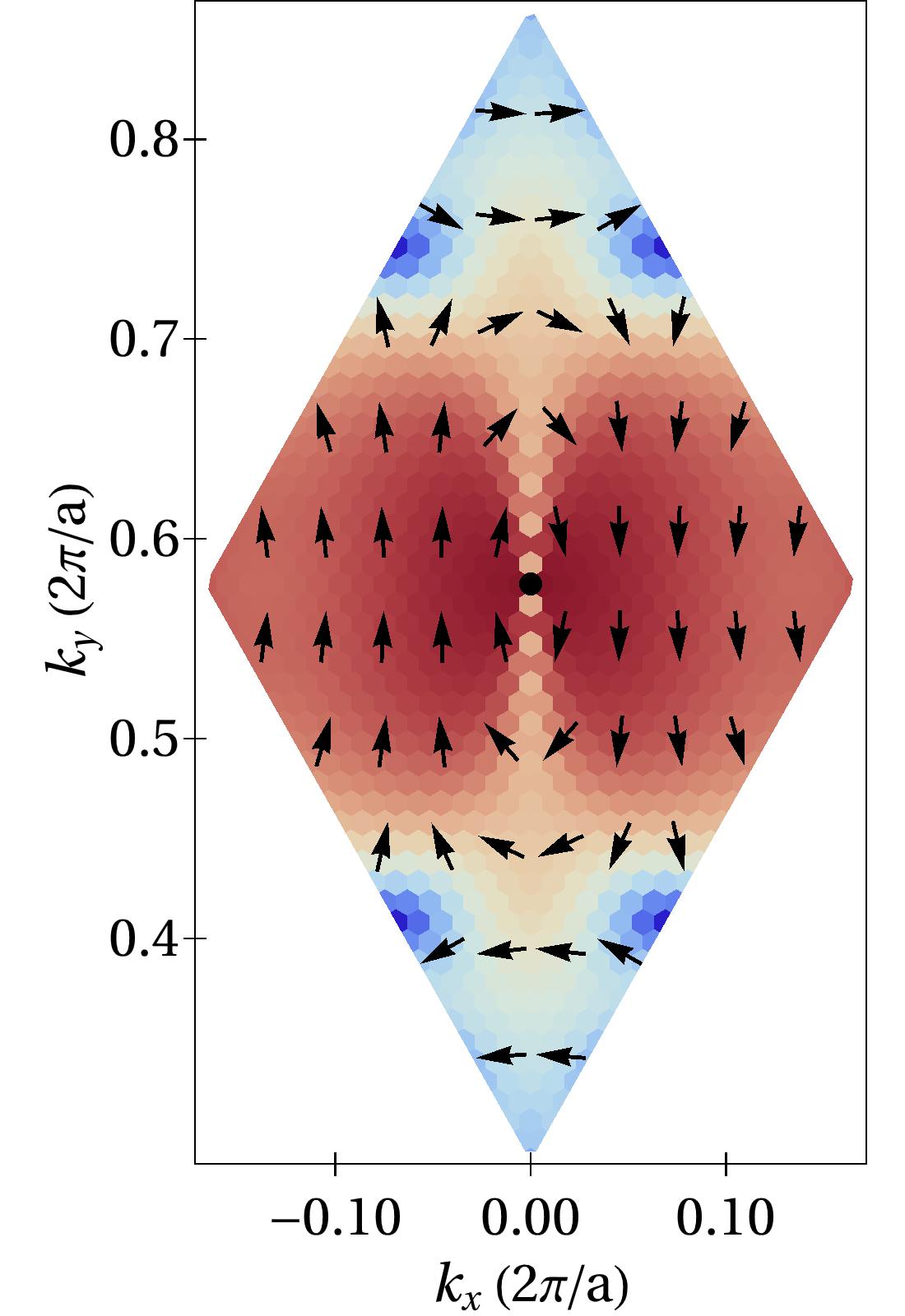} \hspace{0.1cm}\includegraphics[width=0.40\columnwidth]{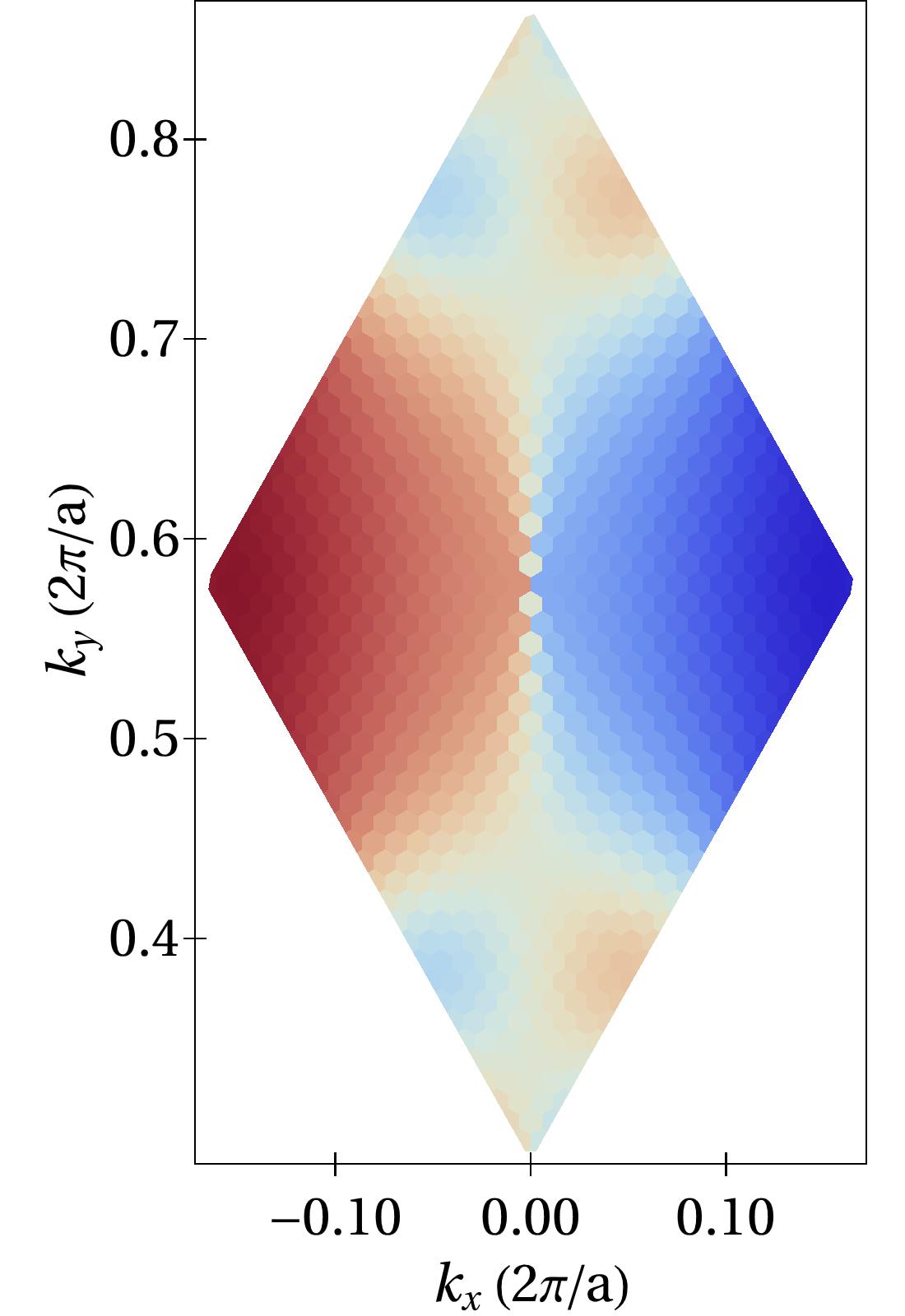} 
\caption{Calculated spin polarization of the upper branch (upper panels) and lower branch (lower panels) of the $\ov{M}$ point Dirac cone of the deep surface states on the Ag(111) surface.  The left panels plot the in-plane spin polarization $(S_x,S_y)$ with the background color showing its modulus (red = maximal, blue = minimal).  The right panels show $S_z$ as a color density plot (red = positive, blue = negative); $S_z/\sqrt{S_x^2+S_y^2}$ ranges between $(-0.65,+0.65)$ for the upper branch and $(-0.75,+0.75)$ for the lower branch.  The plotted region of the surface Brillouin zone is shown in red in the inset.}
\label{fig:Ag:spin:polarization}
\end{figure}

The Ag Dirac cone is closer to the lower edge of the surface-projected bulk band gap, in contrast to Au, where it was near the center of the gap.  The surface states have limited extent in the $\ov{M}\rightarrow\ov{\Gamma}$ direction, vanishing at $k=0.6$ and 1.8~\AA$^{-1}$~when they meet the ``pinching'' of the upper and lower bulk bands.  Along $\ov{M}\rightarrow\ov{K}$, the upper and lower surface state branches are similarly attenuated when they intersect the bulk bands.  As occurred in Au, the upper branch connects with the bulk band above the gap as $\mathbf{k}$ moves from $\ov{M}$ toward $\ov{K}$.  In line with the greater anisotropy of the transverse and longitudinal Rashba parameters in Ag with respect to Au, the spin polarization of the surface states in Fig.~\ref{fig:Ag:spin:polarization} is more strongly anisotropic.


\section{Topological character of the surface states}
\label{sec:topological}

Dirac cones are a characteristic feature of topological insulators, and it is interesting to ask whether there is a topological invariant associated with these deep surface states in Ag and Au.  Although as metals Ag and Au of course cannot be topological insulators, there are nevertheless some large band gaps between these deep $d$ states.  Indeed, the Dirac cone surface states reside in a spin-orbit induced gap of $>1$~eV.  However, that gap is not maintained throughout the surface Brillouin zone and in fact vanishes by symmetry at the $\ov{\Gamma}$ point.  This means that even if we were able to lower the chemical potential to the level of the Dirac cone without perturbing the band structure or surface morphology, Ag and Au would remain metallic.  Therefore, the Z$_2$ topological invariant \cite{fu2007a,fu2007b} cannot be defined, since the gap is not preserved at all time-reversal invariant momenta of the bulk fcc BZ of Ag and Au.  Specifically, the bulk bands above and below the Dirac cone (the 2nd and 3rd lowest energy bands in Fig.~\ref{fig:Au:bulk}) are degenerate by symmetry at $\Gamma$, one of the eight time-reversal invariant momenta of the bulk BZ.  

One could ask if a nonzero topological invariant would emerge if a gap could be opened, e.g., by lifting this degeneracy by deforming the crystal. We find however that even in that case the strong Z$_2$ topological invariant defined in Ref.~\onlinecite{fu2007b} would remain trivially zero, since all $d$ states are even under inversion.  This is a contraindication to the standard topological classification of these states based on parity symmetry, but it does not exclude the existence of other yet-to-be-discovered topological invariants.  

\section{Conclusions}
\label{sec:conclusions}

We have presented a joint theoretical/experimental study of Rashba-type spin splitting at Dirac cones of the deep $d$ orbital surface states of Ag(111) and Au(111), highlighting qualitative differences with the exhaustively studied $sp$-derived $L$-gap Shockley surface states.  The Dirac cone predicted at the $\ov{M}$ point in our \textit{ab initio} surface state DFT calculations is fully verified by accurate ARPES measurements along high symmetry planes and constant energy slices.  The predicted spin splitting and spin polarization at this $\ov{M}$ point Dirac cone are strongly anisotropic and the surface states display a corresponding rhombic warping as opposed to the hexagonal warping at $\ov{\Gamma}$ point Dirac cones.  The anisotropy has been explained in terms of the lower symmetry of the small group of the wave vector at the $\ov{M}$ point.  Most of the other deep surface states are predicted to be strongly spin polarized as a consequence of the large spin-orbit interaction acting on the $d$ manifold, and it would be interesting to verify the predicted spin polarization by spin- and angle-resolved photoemission spectroscopy.

The spin polarization of the observed surface states is probably not directly relevant to spintronics applications, given that the states are buried 2--8~eV below the Fermi energy.  Nevertheless, the deep surface states of Au and Ag present an ideal prototype of $d$ as opposed to $sp$-derived surface states.  Besides exhibiting stronger spin-orbit effects, the $d$-orbital surface states display more complicated spin textures, band dispersion and avoided crossings among the five $d$ subbands throughout the BZ, indicative of multiband spin-orbit interactions, which cannot be described by the Rashba model.  Our theoretical analysis demonstrates that spin-orbit interactions in $d$ orbital Tamm surface states generally have a multiband character, which might have implications in the search for novel spintronics materials and topological insulators.

\begin{acknowledgments}
R.R. acknowledges A.~Dal Corso for help with spin-orbit interactions in \textsf{QuantumEspresso}.  R.R. and work at SISSA have been supported by PRIN-COFIN contract 2010LLKJBX 004.  P.M.S., P.M., and C.C. acknowledge the Progetto Premiale, Materiali e Disposivi Magnetici e Superconduttivi per Sensoristica e ICT of the Italian Ministry of Education, University and Research (MIUR).  S.K.M. was supported by the ICTP-TRIL program, Trieste, Italy.
\end{acknowledgments}

\bibliography{bibliography}

\end{document}